\documentclass[3p,times,twocolumn]{elsarticle}
\pdfoutput=1
\usepackage{ecrc}

\volume{00}

\firstpage{1}

\journalname{Nuclear Physics B Proceedings Supplement}

\runauth{}

\jid{nuphbp}

\jnltitlelogo{Nuclear Physics B Proceedings Supplement}

\usepackage{amssymb}

\usepackage[figuresright]{rotating}
\newcommand{\amu}{$a_{\mu}$}
\begin{document}

\begin{frontmatter}



\dochead{}

\title{The New Muon g-2 experiment at Fermilab}


\author{Graziano Venanzoni}
\ead{graziano.venanzoni@lnf.infn.it}
\author{\\on behalf of the Fermilab E989 collaboration}
\address{Laboratori Nazionali di Frascati dell'INFN, Frascati, Italy}


\begin{abstract}
There is a long standing discrepancy between the Standard Model
prediction for the muon g-2 and the value measured by the Brookhaven
E821 Experiment.  At present the discrepancy stands at about three
standard deviations, with a comparable accuracy between experiment and
theory. Two new proposals -- at Fermilab and J-PARC -- plan to improve the
experimental uncertainty by a factor of 4, and it is expected that there
will be a significant reduction in the uncertainty of the Standard Model
prediction.  I will review the status of the planned experiment at
Fermilab, E989, which will analyse 21 times more muons than the BNL
experiment and discuss how the systematic uncertainty will be reduced
by a factor of 3 such that a precision of 0.14~ppm can be achieved.
\end{abstract}




\end{frontmatter}


\section{Introduction}
\label{Introduction}
\begin{figure}[htb]
\begin{center}
\includegraphics[width=8cm]{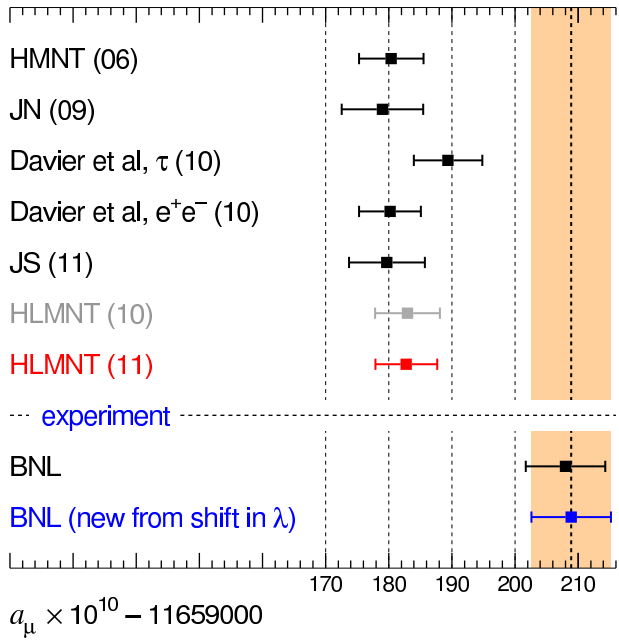}
\vspace{-0.4 cm}
\caption{Standard model predictions of $a_\mu$ by several groups
  compared to the measurement from BNL (taken from~\cite{teubner}).}
\label{fig1}
\end{center}
\end{figure}

The muon anomaly $a_{\mu}=(g-2)/2$ is a low-energy observable, which
can be both measured and computed to high
precision~\cite{Jegerlehner:2008zza,Blum:2013xva}. Therefore it provides an important test
of the Standard Model (SM) and it is a sensitive search for new
physics~\cite{Stockinger:1900zz}.  Since the first precision
measurement of $a_{\mu}$ from the E821 experiment at BNL in
2001~\cite{Brown:2001mga}, there has been a discrepancy between its
experimental value and the SM prediction.  The significance of this discrepancy
has been slowly growing due to reductions in the theory uncertainty.
Figure~\ref{fig1} (taken from~\cite{teubner}) shows a recent
comparison of the SM predictions of different groups and the BNL
measurement for $a_{\mu}$.  The $a_{\mu}$ determinations of the
different groups are in very good agreement and show a consistent
$\approx 3\,\sigma$ discrepancy~\cite{teubner,Jegerlehner:2009ry,Davier:2010nc},
despite many recent iterations in the SM calculation.  It should be
noted that with the final E821 measurement and advances in the 
theoretical SM calculation that both the theory and experiment
uncertainties have been reduced by more than a factor two in the last
ten years~\cite{Prades:2009qp}.
The accuracy of the theoretical prediction ($\delta
a_{\mu}^{\rm{TH}}$, between 5 and 6 $\times 10^{-10}$) is limited by
the strong interaction effects which cannot be computed perturbatively
at low energies. The leading-order hadronic vacuum polarization
contribution, $a_{\mu}^{\rm{HLO}}$, gives the main uncertainty
(between 4 and 5 $\times 10^{-10}$).  It can be related by a dispersion
integral to the measured hadronic cross sections, and it is known with
a fractional accuracy of 0.7\%, i.e. to about 0.4 ppm.  The
O($\alpha^3$) hadronic light-by-light contribution,
$a_{\mu}^{\rm{HLbL}}$, is the second dominant error in the theoretical
evaluation.  It cannot at present be determined from data, and relies
on using specific models.  Although its value is almost two orders of
magnitude smaller than $a_{\mu}^{\rm{HLO}}$, it is much worse known
(with a fractional error of the order of 30\%) and therefore it still
give a significant contribution to $\delta a_{\mu}^{\rm{TH}}$
(between 2.5 and 4 $\times 10^{-10}$).\\ 

From the experimental side, the error achieved by the BNL E821
experiment is $\delta a_{\mu}^{\rm{EXP}}= 6.3 \times 10^{-10}$ (0.54
ppm)~\cite{Bennett:2006fi}.  This impressive result is still limited
by the statistical errors, and a new experiment, E989~\cite{g-2}, to
measure the muon anomaly to a precision of $1.6 \times 10^{-10}$ (0.14
ppm) is under construction at Fermilab. If the central value remains
unchanged, then the statistical significance of the discrepancy with
respect to the SM prediction would then be over 5$\sigma$, see
Ref.~\cite{Blum:2013xva}, and would be larger than this with the
expected improvements in the theoretical calculation.


\section{Recent results and expected improvement on the hadronic contribution}
In contrast to the QED and Electroweak contributions to \amu, which
can be calculated using perturbation theory, and therefore are well
under control, the hadronic contributions (LO VP and HLbL) cannot be computed
reliably using perturbative QCD.  The hadronic contribution
$a_{\mu}^{\rm{HLO}}$ can can be computed from hadronic $e^+ e^-$
annihilation data via a dispersion relation, and therefore its
uncertainty strongly depends on the accuracy of the experimental
data. For the Hadronic Light-by-Light contribution
$a_{\mu}^{\rm{HLbL}}$ there is no direct connection with data and
therefore only model-dependent estimates exist.  As the hadronic
sector dominates the uncertainty on the theoretical prediction
$a_{\mu}^{\rm{TH}}$, it has been the subject of considerable recent activity from both 
experimental and theoretical groups, with the following outcomes:
\begin{itemize}
\item A precise determination of the hadronic cross sections at the
  $e^+e^-$ colliders (VEPP-2M, DA$\mathrm{\Phi}$NE, BEPC, PEP-II and
  KEKB) has allowed a determination of $a_{\mu}^{\rm{HLO}}$ with a
  fractional accuracy below 1\%.  These efforts have led to the development
  of dedicated high precision theoretical tools such as the addition of Radiative
  Corrections (RC) and the non-perturbative hadronic contribution to
  the running of $\alpha$ (i.e. the vacuum polarisation, VP) into the Monte
  Carlo (MC) programs used for the analysis of the
  experimental data~\cite{Actis:2010gg};

\item The use of `\emph{Initial State Radiation}' (ISR) data~\cite{Chen:1974wv,Binner:1999bt,Benayoun:1999hm} which has opened a
  new way to precisely obtain the electron-positron annihilation cross
  sections into hadrons at particle factories operating at fixed
  beam-energies~\cite{Kluge:2008fb,Druzhinin:2011qd}.

\item A dedicated effort on the evaluation of the Hadronic
  Light-by-Light contribution (see for example ~\cite{int}), where two
  different groups~\cite{Prades:2009tw,Jegerlehner:2009ry} have
  obtained consistent values (with slightly different errors), and
  therefore strengthened the confidence in the reliability of these
  estimates.

\item Impressive progress on the lattice, where 
an accuracy of $\sim 4\%$ has been reached 
on the four-flavour calculation of $a_{\mu}^{\rm{HLO}}$~\cite{Burger:2013jya};

\item A better agreement between the $e^+e^- -$ and the $\tau-$ based
  evaluation of $a_{\mu}^{\rm{HLO}}$, due to improved isospin
  corrections~\cite{Davier:2010nc}.  These two sets of data are
  now broadly in agreement (with $\tau$ data moving towards $e^+e^- -$
  data) after including vector meson and $\rho - \gamma$ mixing
  \cite{Jegerlehner:2011ti,Benayoun:2012wc}.
\end{itemize}

Further improvements are expected on the calculations of the hadronic
contribution to \amu\ on the timescale of the new g-2 experiments at
Fermilab and J-PARC and this will be augmented, on the experimental side, by more data 
from current and future $e^+ e^-$ colliders.  From the theoretical side, the
lattice calculation has already reached a mature stage and has real
prospects to match the experimental precision. From both
activities a further reduction of the error on $a_{\mu}^{\rm{HLO}}$
can be expected and thus progress on  $a_{\mu}^{\rm{HLbL}}$ will be required.
Although for the HLbL contribution there isn't a direct connection with data,
 $\gamma-\gamma$ measurements performed at $e^+ e^-$ colliders will help 
constrain the on-shell form factors~\cite{Babusci:2011bg,Colangelo:2014pva} and 
lattice calculations will help better determine the off shell contributions.

\section{Measuring \amu}
The measurement of \amu\ uses the spin precession
resulting from the torque experienced by
the magnetic moment when placed in a magnetic
field. An ensemble of polarized muons is introduced
into a magnetic field, where they are stored
for the measurement period. With the assumption that the
muon velocity is transverse to the magnetic field
($\vec{\beta}\cdot\vec{B} = 0$), the rate at which the spin turns
relative to the momentum vector is given by the
difference frequency between the spin precession
and cyclotron frequencies. 
Because electric quadrupoles are used to provide vertical focusing in the storage ring, their
electric field is seen in the muon rest frame as a motional magnetic field that can affect the
spin precession frequency.
In the presence of both $\vec{E}$ 
and $\vec{B}$ fields, and in the case that $\vec{\beta}$
is perpendicular to both, the anomalous precession frequency ({\it i.e.} the frequency at which 
the muon’s spin advances relative to its
momentum)
is 
\begin{eqnarray}
\nonumber
\vec{\omega_a} & =& \vec{\omega_S} - \vec{\omega_C} \\
 & = & -\frac{q}{m}\Big [a_{\mu}\vec{B} - \Big (a_{\mu} - \frac{1}{\gamma^2-1}\Big )\frac{\vec{\beta}\times \vec{E}}{c}\Big ]\label{eq1}
\end{eqnarray}

The experimentally measured
numbers are the muon spin frequency $\omega_a$ and the
magnetic field, which is measured with proton
NMR, calibrated to the Larmor precession 
frequency, $\omega_p$, of a free proton. The anomaly is related
to these two frequencies by
\begin{eqnarray}
a_{\mu} & = & \frac{\tilde{\omega_a}/\omega_p}{\lambda-\tilde{\omega_a}/\omega_p} = 
\frac{R}{\lambda R},
\end{eqnarray}
where $\lambda = \mu_\mu/\mu_p = 3.183 345 137(85)$ (determined
experimentally from the hyperfine structure of muonium), and $ R =
\tilde{\omega_a}/\omega_p$ .  The tilde over $\omega_a$ means it has
been corrected for the spread in the beam momentum (the so-called
electric-field correction) and for the vertical betatron oscillations
which mean that $\vec{\beta}\cdot\vec{B} \neq$ (the so-called pitch
corrections): these are the only corrections made to the measurement.
The magnetic field in Eq.~(\ref{eq1})
is an average that can be expressed as an integral
of the product of the muon distribution times
the magnetic field distribution over the storage
region. Since the moments of the muon distribution
couple to the respective multipoles of the
magnetic field, either one needs an exceedingly
uniform magnetic field, or exceptionally good information
on the muon orbits in the storage ring,
to determine $<B>_{\mu-dist}$ to sub-ppm precision.
This was possible in E821 where the
uncertainty on the magnetic field averaged over
the muon distribution was 30 ppb (parts per
billion). 
The coefficient of the $\vec{\beta}\times\vec{E}$ 
term in Eq.~(\ref{eq1}) vanishes at the ``magic" momentum of 3.094 GeV/c
where $\gamma = 29.3$. Thus \amu\ can be determined by a precision measurement of $\omega_a$ and B. 
At this magic momentum, the electric field  is used only for muon storage and the magnetic 
field alone determines the precession frequency. The finite spread in beam momentum
and vertical betatron oscillations introduce small (sub ppm) corrections to the precession
frequency. These are the only corrections made to the measurement.

The experiment consists of repeated fills of the
storage ring, each one introducing an ensemble
of muons into a magnetic storage ring, and then
measuring the two frequencies $\omega_a$ and $\omega_p$. The
muon lifetime is 64.4 $\mu$s, and the
data collection period is typically  700 $\mu$s. The g-2 precession period 
is 4.37 $\mu$s, and the cyclotron period $\omega_C$ is 149 ns.

\begin{figure}[htb]
\begin{center}
\includegraphics[width=8cm,angle=0]{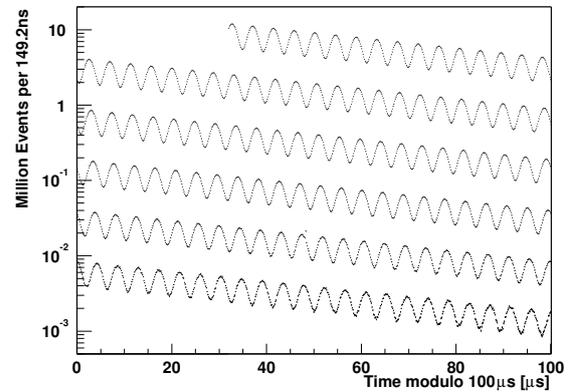}
\vspace{-.5cm}
\caption{Distribution of electron counts versus time for 3.6 billion
  muon decays from the E821 experiment. The data are wrapped around
  modulo 100 $\mu$s~\cite{Bennett:2006fi}.}
\label{fig2}
\end{center}
\end{figure}

Because of parity violation in the weak decay of the muon, a correlation exists between
the muon spin and the direction of the high-energy decay electrons.
Thus as the spin turns relative to the momentum,
the number of high-energy decay electrons
is modulated by the frequency $\omega_a$, as shown
in Fig.~\ref{fig2}.
The E821 storage ring was constructed as a
“super-ferric” magnet, meaning that the iron
determined the shape of the magnetic field. Thus
$B_0$ needed to be well below saturation and was
chosen to be 1.45 T. The resulting ring had a 
central orbit radius of 7.112 m, and 24 detector stations
were placed symmetrically around the inner
radius of the storage ring. 
The detectors were made of Pb/SciFi electromagnetic calorimeters 
which measured the decay electron energy and time of arrival.
The detector geometry
and number were optimized to detect the high energy
decay electrons, which carry the largest
asymmetry, and thus information on the muon
spin direction at the time of decay. In this design
many of the lower-energy electrons miss the
detectors, reducing background and pileup.


\section{The FERMILAB PROPOSAL: E989}
The E989 experiment at Fermilab plans to measure \amu\ to an uncertainty of $16\times 10^{11}$ (0.14 ppm), 
derived from a 0.10 ppm
statistical error and roughly equal 0.07 ppm systematic uncertainties on $\omega_a$ and $\omega_p$.


The proposal efficiently uses the unique properties of the Fermilab
beam complex to produce the necessary flux of muons, which will be
injected and stored in the (relocated) muon storage ring.  To achieve
a statistical uncertainty of 0.1 ppm, the total data set must contain
more than $1.8\times 10^{11}$ detected positrons with energy greater
than 1.8 GeV, and arrival time greater than 30 $\mu$s after injection
into the storage ring. Four out of 20 of the 8-GeV Booster proton
batches in 15 Hz operational mode, each subdivided into four bunches
of intensity $10^{12}$ p/bunch, will be used to provide muons. The
proton bunches fill the muon storage ring at a repetition rate of 12
Hz, to be compared to the 4.4 Hz at BNL.  The proton bunch hits a
target in the antiproton area, producing a 3.1 GeV/c pion beam that is
directed along a nearly 2000 m decay line, including several revolutions around the Delivery Ring, which are used to further eliminate pions and to displace secondary protons from muons using time of flight and a kicker to sweep out the protons.  The resulting pure muon beam is
injected into the storage ring.  The muons enter the ring through a
superconducting inflector magnet. At present it is envisaged that the
BNL inflector will be used but there is a vigorous R\&D programme
underway investigating the possible use of a new large aperture 
inflector that would increase the number of stored muons and reduce
the multiple scattering.
A better optimized pulse-forming network will energize the storage
ring kicker to place the beam on a stable orbit.
The pion flash (caused by pions entering the ring at injection) will
be eliminated owing to the long beamline,
and the muon flux
will be significantly increased because of the ability to take
zero-degree muons. 


\begin{figure}[h]
\begin{center}
\includegraphics[width=8cm,angle=0]{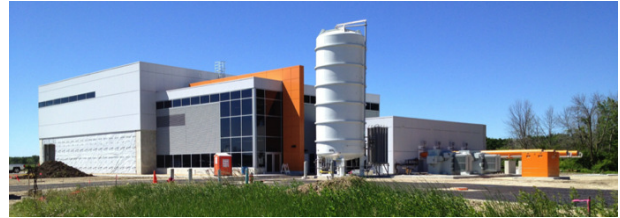}
\vspace{-.5cm}
\caption{The new MC-1 building at Fermilab, where the muon g-2 storage ring is being reassembled in the larger part to the left. The
part to the right houses the counting room, electronics, etc, with cryogenics services further right. (Image credit: Fermilab.)}
\label{fig3}
\end{center}
\end{figure}
\begin{figure}[t]
\begin{center}
\includegraphics[width=8cm,angle=0]{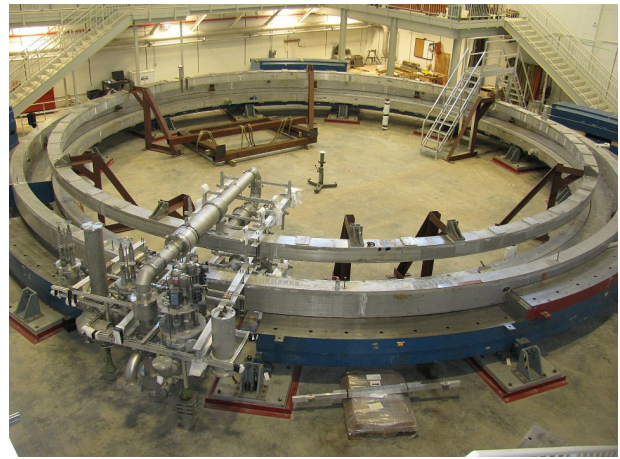}
\vspace{-.5cm}
\caption{Re-assembly of the g-2 storage-ring magnet at Fermilab,
after the three superconducting coils were positioned gently on top
of the newly assembled bottom ring of steel yoke segments. The
coils and their complex interconnect system (top right in photo)
were transported as a single unit from Brookhaven to Fermilab by
land, sea and river, in 2013. (Image credit: Fermilab.)}
\label{fig4}
\end{center}
\end{figure}

In the summer of 2013 the E821 muon storage has been moved from
Brookhaven to Fermilab and it has been already relocated in the newly
completed MC-1 building at Fermilab (see Figs.~\ref{fig3} and
~\ref{fig4}) with a stable floor and good temperature control, neither
of which were available at Brookhaven.


The new experiment will require upgrades of detectors, electronics and data acquisition
equipment to handle the much higher data volumes and slightly higher instantaneous rates. Electromagnetic calorimeters made of lead fluoride (PbF2) crystals, with
large area (1.2$\times$1.2 cm) Silicon Photo-Multiplier (SiPM) readout will be used.
A prototype matrix made of 28 crystals together with SiPM bias power supply and a laser diode based monitoring system has been successfully tested  at the SLAC National Accelerator Laboratory (SLAC) test beam facility~\cite{nim}. 
In-vacuum straw drift tubes will be used to measure the characteristics of the muon beam,
and provide data for an improved muon electric dipole moment
measurement, which can be obtained in parallel~\cite{Bennett:2008dy}.
A modern data acquisition system will be used to read out waveform
digitizer data and store it so that both the traditional event mode
and a new integrating mode of data analysis can both be used in
parallel.  The systematic uncertainty on the precession frequency is
expected to improve by a factor 3 thanks to the reduced pion
contamination, the segmented detectors, and an improved storage ring
kick of the muons onto orbit.  The storage ring magnetic field will be
shimmed to an even more impressive uniformity, and improvements in the
field-measuring system will be implemented.  The systematic error on
the magnetic field is halved by better shimming, 
and other incremental changes.  In less than two
years of running, the statistical goal of $4\times 10^{20}$ protons on
target can be achieved for positive muons. 
A follow-up measurement using negative muons is possible.
Two additional physics results will be obtained from the
same data: a new limit on the muon's electric dipole moment; 
and, a more stringent limit on possible CPT or Lorentz
violation in muon spin precession.  The first physics data-taking is
expected in early 2017. The next critical milestone will be the
cooling of the superconducting coils and the powering of the storage-ring
magnet, which is expected by the spring of 2015.

\section{Conclusion}
The measurements of the muon g-2 have been an important benchmark
 for the development of QED and the Standard Model. 
In the recent years, following the impressed accuracy (0.54 ppm) reached by E821 experiment at BNL,
 a worldwide effort from different theoretical and experimental groups 
have significant improved its SM prediction.
At present 
there appears to be a 3$\sigma$ difference between the theoretical (SM) and the experimental value.
This discrepancy, which  would fit well with SUSY expectations and other beyond the Standard Model theories, is a 
valuable constraint in restricting physics beyond the standard model and  guiding  
the interpretation of  LHC results.
In order to clarify the nature of the observed discrepancy between theory and experiment
and eventually firmly establish (or constrain) new physics effects,
new direct measurements of the
muon g-2 with a fourfold improvement in accuracy have been proposed at Fermilab by E989 experiment, and J-PARC.
First results from E989 are expected around 2017/18.


\begin{thebibliography}{00}


\bibitem{Jegerlehner:2008zza}
  F.~Jegerlehner,
  ``The anomalous magnetic moment of the muon,''
  Springer Tracts Mod.\ Phys.\  {\bf 226} (2008) 1

\bibitem{Blum:2013xva}
  T.~Blum, A.~Denig, I.~Logashenko, E.~de Rafael, B.~Lee Roberts, T.~Teubner and G.~Venanzoni,
  arXiv:1311.2198 [hep-ph]

\bibitem{Stockinger:1900zz}
  D.~St\"ockinger,
 ``Muon (g-2) and physics beyond the standard model,'' In Roberts, Lee B., Marciano, William J. (eds.): Lepton dipole moments 393-438
  (Advanced series on directions in high energy physics. 20)


\bibitem{Brown:2001mga}
  H.~N.~Brown {\it et al.}  [Muon g-2 Collaboration],
  Phys.\ Rev.\ Lett.\  {\bf 86} (2001) 2227

\bibitem{teubner}
  K.~Hagiwara, R.~Liao, A.~D.~Martin, D.~Nomura and T.~Teubner,
  J.\ Phys.\ G {\bf 38} (2011) 085003

\bibitem{Jegerlehner:2009ry}
  F.~Jegerlehner and A.~Nyffeler,
  Phys.\ Rept.\  {\bf 477} (2009) 1
\bibitem{Davier:2010nc}
 M.~Davier, A.~Hoecker, B.~Malaescu and Z.~Zhang,
  Eur.\ Phys.\ J.\ C {\bf 71} (2011) 1515
   [Erratum-ibid.\ C {\bf 72} (2012) 1874]




\bibitem{Prades:2009qp}
  J.~Prades,
  Acta Phys.\ Polon.\ Supp.\  {\bf 3} (2010) 75


\bibitem{Bennett:2006fi}
  G.~W.~Bennett {\it et al.}  [Muon G-2 Collaboration],
  Phys.\ Rev.\ D {\bf 73} (2006) 072003
\bibitem{g-2}
New Muon (g − 2) Collaboration, R.M.~ Carey {\it et. al.}, see
http://lss.fnal.gov/archive/testproposal/0000/fermilab-proposal-0989.shtml


\bibitem{Actis:2010gg}
  S.~Actis {\it et al.}  [Working Group on Radiative Corrections and Monte Carlo Generators for Low Energies Collaboration],
  Eur.\ Phys.\ J.\ C {\bf 66} (2010) 585

\bibitem{Chen:1974wv}
  M.~S.~Chen and P.~M.~Zerwas,
  Phys.\ Rev.\ D {\bf 11} (1975) 58

\bibitem{Binner:1999bt}
  S.~Binner, J.~H.~Kuhn and K.~Melnikov,
  Phys.\ Lett.\ B {\bf 459} (1999) 279

\bibitem{Benayoun:1999hm}
  M.~Benayoun, S.~I.~Eidelman, V.~N.~Ivanchenko and Z.~K.~Silagadze,
  Mod.\ Phys.\ Lett.\ A {\bf 14} (1999) 2605



\bibitem{Kluge:2008fb}
  W.~Kluge,
  Nucl.\ Phys.\ Proc.\ Suppl.\  {\bf 181-182} (2008) 280

\bibitem{Druzhinin:2011qd}
  V.~P.~Druzhinin, S.~I.~Eidelman, S.~I.~Serednyakov and E.~P.~Solodov,
  Rev. Mod. Phys. {\bf 83}, (2011) 1545 


\bibitem{int}
 http://www.int.washington.edu/PROGRAMS/11-47w/

\bibitem{Prades:2009tw}
  J.~Prades, E.~de Rafael and A.~Vainshtein,
  (Advanced series on directions in high energy physics. 20)


\bibitem{Burger:2013jya}
  F.~Burger {\it et al.}  [ETM Collaboration],
  JHEP {\bf 1402} (2014) 099
 

\bibitem{Jegerlehner:2011ti}
  F.~Jegerlehner and R.~Szafron,
  Eur.\ Phys.\ J.\ C {\bf 71} (2011) 1632

\bibitem{Benayoun:2012wc}
  M.~Benayoun, P.~David, L.~DelBuono and F.~Jegerlehner,
  Eur.\ Phys.\ J.\ C {\bf 73} (2013) 2453

\bibitem{Babusci:2011bg}
  D.~Babusci,  {\it et al.},
  Eur.\ Phys.\ J.\ C {\bf 72} (2012) 1917

\bibitem{Colangelo:2014pva}
  G.~Colangelo, M.~Hoferichter, B.~Kubis, M.~Procura and P.~Stoffer,
  Phys.\ Lett.\ B {\bf 738} (2014) 6


\bibitem{nim}
A.~Fienberg {\it et al.},  Nucl.\ Instrum.\ Meth.\ {\it in preparation}



\bibitem{Bennett:2008dy}
  G.~W.~Bennett {\it et al.}  [Muon (g-2) Collaboration],
  Phys.\ Rev.\ D {\bf 80} (2009) 052008

\end{thebibliography}
\end{document}